\documentstyle[12pt,epsf,amsfonts,amssymb,amsbsy]{article}
\newcommand\VV{\setbox0=\hbox{V}\hbox{\rm V\raise\ht0
  \hbox to0pt{\hss\vbox to0pt{\hbox{v}\vss}}}}
\def\slashchar#1{\setbox0=\hbox{$#1$}           % set a box for #1
   \dimen0=\wd0                                 % and get its size
   \setbox1=\hbox{/} \dimen1=\wd1               % get size of /
   \ifdim\dimen0>\dimen1                        % #1 is bigger
      \rlap{\hbox to \dimen0{\hfil/\hfil}}      % so center / in box
      #1                                        % and print #1
   \else                                        % / is bigger
      \rlap{\hbox to \dimen1{\hfil$#1$\hfil}}   % so center #1
      /                                         % and print /
   \fi}                                         %
\begin{document}
\vspace*{3cm}
\begin{center}
{\large \bf Lifetimes of doubly heavy baryons}\\

\vspace*{5mm}
A.K. Likhoded$^{a)}$, A.I.
Onishchenko$^{b)}$\\

\vspace*{3mm}
{a) \sf State Research Center of Russia
"Institute for High Energy Physics"} \\ {\it Protvino, Moscow
region, 142284 Russia}\\ {b) \sf Institute for Theoretical and
Experimental Physics}\\ {\it Moscow, B. Cheremushkinskaja, 25, 117259
Russia \\
Fax: 7 (095) 123-65-84}
\end{center}

\begin{abstract}
We perform a detailed investigation of total lifetimes for the doubly heavy
baryons $\Xi_{QQ'}$, $\Omega_{QQ'}$ in the framework of operator product
expansion over the inverse heavy quark mass, whereas, to estimate matrix
elements of operators obtained in OPE, approximations of nonrelativistic QCD
are used.
\end{abstract}

%\vspace*{2.cm}

\section{Introduction}

At present a number of powerful techniques based on Operator Product Expansion
(OPE) and effective field theories have been developed. These tools allow one
consistently to include into consideration various nonperturbative
contributions, written in terms of a few number of universal quantities. The
coefficients (Wilson coefficients) in front of these operators are generally
expanded in series over the QCD coupling constant, inverse heavy quark mass
and/or relative velocity of heavy quarks inside the hadron. The accuracy,
obtained in such calculations, can be systematically improved, and it is
limited only by the convergence properties of the mentioned series. The
described approach have been already widely used for making the precise
predictions in the heavy quark sector of Standard Model (SM), such as decays,
distributions and partial width asymmetries involving the CP
violation\footnote{For review see \cite{al}.} for the heavy hadrons. The
sensitivity of Wilson coefficients to virtual corrections caused by some
higher-scale interactions makes this approach to be invaluable in searching for
a "new" physics at forthcoming experiments.

The approach under discussion has been successfully used in the
description of weak decays of the hadrons containing a single heavy
quark, as carried out in the framework of Heavy Quark Effective
Theory (HQET) \cite{HQET}, in the annihilation and radiative
decays of heavy quarkonia $Q\bar Q$, where one used the framework of
non-relativistic QCD (NRQCD) \cite{NRQCD}, and in the weak decays
of long-lived heavy quarkonium with mixed flavours $B_c^+$
\cite{Beneke} \footnote{The first experimental observation of the
$B_c$-meson was recently reported by the CDF Collaboration
\cite{CDFbc}; see ref.\cite{thbc} for a theoretical review of $B_c$-meson
physics before the observation.}. The experimental data on the weak
decays of heavy hadrons can be used for the determination of basic
properties of weak interactions at a fundamental level, in
particular, for the extraction of CKM matrix elements. The same approach is
also valid for the baryons containing two heavy quarks.

In addition to the information extracted from the analysis of hadrons with a
single heavy flavor, the baryons with two heavy quarks, $QQ^\prime q$, provide
a way to explore the nonspectator effects, where their importance is increased.
Here we would like to note, that in the case of systems with two heavy quarks,
the hypothesis on the quark-hadron  duality is more justified, and, so, the
results of OPE-based approach turn out to be more reliable. For these baryons
we can apply a method, based on the combined HQET-NRQCD techniques
\cite{HQET,NRQCD,Beneke}, if we use the quark-diquark picture for the bound
states. The expansion in the inverse heavy quark mass for the heavy diquark
$QQ^\prime$ is a straightforward generalization of these techniques in the
mesonic decays of $B_c$ \cite{NRQCD,Beneke}, with the difference that, instead
of the color-singlet systems, we deal with the color--anti-triplet ones, with
the appropriate account for the interaction with the light quark. First
estimates of the lifetimes for the doubly heavy baryons $\Xi_{cc}^{\diamond}$
and $\Xi_{bc}^{\diamond}$ were recently performed in \cite{DHD1,DHD2}. Using
the same approach, but different values of parameters\footnote{See comments on
the difference in the numerical values of lifetimes of doubly charmed baryons
in \cite{Onishchenko}.} a repetition of our results for the case of doubly
charmed baryons was done in \cite{Guberina}. The spectroscopic characteristics
of baryons with two heavy quarks and the mechanisms of their production in
different interactions were discussed in refs.~\cite{DHS1,DHS2,DHS3,DHSR} and
\cite{prod}, respectively.

In this paper, we present the calculation of lifetimes for the doubly heavy
baryons as well as reconsider the previous estimates with a use of slightly
different set of parameters adjusted in the consideration of lifetime data for
the observed heavy hadrons and improved spectroscopic inputs. As we made in the
description of inclusive decays of the $\Xi_{cc}^{\diamond}$ and
$\Xi_{bc}^{\diamond}$-baryons, we follow the papers \cite{Beneke,vs}, where all
necessary generalizations to the case of hadrons with two heavy quarks and
other corrections are discussed. We note, that in the leading order of OPE
expansion, the inclusive widths are determined by the mechanism of spectator
decays involving free quarks, wherein the corrections in the perturbative QCD
are taken into account. The introduction of subleading terms in the expansion
over the inverse heavy quark masses\footnote{It was shown in \cite{bigi} that
the first order $1/m_Q$-correction is absent, and the corrections begin with
the $1/m_Q^2$-terms.} allows one to take into account the corrections due to
the quark confinement inside the hadron. Here, an essential role is played by
both the motion of heavy quark inside the hadron and chromomagnetic
interactions of quarks. The important ingredient of such corrections in the
baryons with two heavy quarks is the presence of a compact heavy diquark, which
implies that the square of heavy quark momentum is enhanced in comparison with
the corresponding value for the  hadrons with a single heavy quark. The next
characteristic feature of baryons with two heavy quarks is the significant
numerical impact on the lifetimes by the quark contents of hadrons, since in
the third order over the inverse heavy quark mass, $1/m_Q^3$,  the four-quark
correlations in the total width are enforced in the effective lagrangian due to
the two-particle phase space of intermediate states (see the discussion in
\cite{vs}). In this situation, we have to add the effects of Pauli interference
between the products of heavy quark decays and the quarks in the initial state
as well as the weak scattering involving the quarks composing the hadron. Due
to such terms we introduce the corrections depending on spectators and
involving the masses of light and strange quarks in the framework of
non-relativistic models with the constituent quarks, because they determine the
effective physical phase spaces, strongly deviating from the naive estimates in
th decays of charmed quarks. We take into account the corrections to the
effective weak lagrangian due to the evolution of Wilson coefficients from the
scale of the order of heavy quark mass to the energy, characterizing the
binding of quarks inside the hadron \cite{vs}.

The paper is organized as follows. In agreement with the general picture given
above, we describe the scheme for the construction of OPE
for the total width of baryons containing two heavy quarks with account
of corrections to the spectator widths in Section 2. The procedure for the
estimation of non-perturbative matrix elements of operators in the doubly heavy
baryons is considered in Section 3 in terms of non-relativistic heavy quarks.
Section 4 is devoted to the numerical evaluation and discussion of parameter
dependence of lifetimes of doubly heavy baryons.  We conclude in
section 5 by summarizing our results.

\section{Description of the method}
In this section we describe the approach used for the calculation of total
lifetimes for the doubly heavy baryons, originally formulated in
\cite{DHD1,DHD2}, together with some new formulae, required for the evaluation
of nonspectator effects in the decays of other\footnote{Others mean those of
not considered in \cite{DHD1,DHD2}.} baryons in the family of doubly heavy
baryons, not considered previously.

The optical theorem along with the hypothesis of integral quark-hadron duality,
leads us to a relation between the total decay width of heavy quark and the
imaginary part of its forward scattering amplitude. This relationship, applied
to the $\Xi_{QQ'}^{(*)}$-baryon total decay width $\Gamma_{\Xi_{QQ'}^{(*)}}$,
can be written down as:
\begin{equation}
\Gamma_{\Xi_{QQ'}^{(*)}} =
\frac{1}{2M_{\Xi_{QQ'}^{(*)}}}\langle\Xi_{QQ'}^{(*)}|{\cal T}
|\Xi_{QQ'}^{(*)}\rangle ,
\label{1}
\end{equation}
where the $\Xi_{QQ'}^{(*)}$ state in Eq. (\ref{1}) has the ordinary
relativistic
normalization, $\langle \Xi_{QQ'}^{(*)}|\Xi_{QQ'}^{(*)}\rangle  = 2EV$, and
the
transition operator ${\cal T}$ is determined by the expression
\begin{equation}
{\cal T} = \Im m\int d^4x~\{{\hat T}H_{eff}(x)H_{eff}(0)\},
\end{equation}
where $H_{eff}$ is the standard effective hamiltonian, describing the low
energy interactions of initial quarks with the decays products, so that
\begin{equation}
H_{eff} = \frac{G_F}{2\sqrt 2}V_{q_3q_4}V_{q_1q_2}^{*}[C_{+}(\mu)O_{+} +
C_{-}(\mu)O_{-}] + h.c.
\end{equation}
where
$$
O_{\pm} = [\bar q_{1\alpha}\gamma_{\nu}(1-\gamma_5)q_{2\beta}][\bar
q_{3\gamma}\gamma^{\nu}(1-\gamma_5)q_{4\delta}](\delta_{\alpha\beta}\delta_{
\gamma\delta}\pm\delta_{\alpha\delta}\delta_{\gamma\beta}),
$$
and
$$
C_+ = \left [\frac{\alpha_s(M_W)}{\alpha_s(\mu)}\right ]^{\frac{6}{33-2f}},
\quad
C_- = \left [\frac{\alpha_s(M_W)}{\alpha_s(\mu)}\right ]^{\frac{-12}{33-2f}},\\
$$
where f is the number of flavors, $\{\alpha,\beta,\gamma,\delta \}$ run over
the
color indeces.

Under an assumption, that the energy release in the heavy quark decay is large,
we can perform the operator product expansion for the transition operator
${\cal T}$ in Eq.(\ref{1}). In this way we obtain series of local operators
with increasing dimensions over the energy scale, wherein the contributions to
$\Gamma_{\Xi_{QQ'}^{(*)}}$ are suppressed by the increasing inverse powers of
the heavy quark masses. This formalism has already been applied to calculate
the total decay rates for the hadrons, containing a single heavy quark
\cite{bigi} (for the most early work, having used similar methods, see also
\cite{vs,GRT}) and hadrons, containing two heavy quarks \cite{DHD1,DHD2}. As
was already pointed in \cite{DHD1}, the expansion, applied here, is
simultaneously in the powers of both inverse heavy quark masses and the
relative velocity of heavy quarks inside the hadron. Thus, this fact shows the
difference between the description for the doubly heavy baryons and the
consideration of both the heavy-light mesons (the expansion in powers of
$\frac{\Lambda_{QCD}}{m_Q}$) and the heavy-heavy mesons \cite{Beneke} (the
expansion in powers of relative velocity of heavy quarks inside the hadron,
where one can apply the scaling rules of nonrelativistic QCD \cite{4}).

The operator product expansion explored has the form:
\begin{equation}
{\cal T} = \sum_{i=1}^2 C_1(\mu)\bar Q^iQ^i + \frac{1}{m_{Q^i}^2}C_2(\mu)\bar
Q^i g\sigma_{\mu\nu}G^{\mu\nu}Q^i
+ \frac{1}{m_{Q^i}^3}O(1) \label{4}
\end{equation}

The leading contribution in Eq.(\ref{4}) is determined by the operators $\bar
Q^iQ^i$,
corresponding to the spectator decay of $Q^i$-quarks. The use of
motion equation for the heavy quark fields allows one to eliminate some
redundant operators, so that no operators of dimension four contribute. There
is a single operator of dimension five, $Q^i_{GQ} = \bar Q^i g \sigma_{\mu\nu}
G^{\mu\nu} Q^i$. As we will show below, significant contributions come from the
operators of dimension six $Q_{2Q^i2q} = \bar Q^i\Gamma q\bar q\Gamma^{'}Q^i$,
representing the effects of Pauli interference and weak scattering for
doubly heavy baryons. Furthermore, there are also other operators of dimension
six
$Q_{61Q^i} = \bar Q^i \sigma_{\mu\nu}\gamma_{l} D^{\mu}G^{\nu l}Q^i$ and
$Q_{62Q^i} = \bar Q^i D_{\mu} G^{\mu\nu}\Gamma_{\nu}Q^i$, which are
suppressed in comparison with $Q_{2Q^i2q}$ \cite{vs}. In what follows, we do
not calculate the corresponding coefficient functions for the latter two
operators, so that the expansion is certainly complete up to the second order
of $\frac {1}{m}$, only.

Further, the different contributions to OPE are given by the following:
\begin{eqnarray}
{\cal T}_{\Xi_{cc}^{++}} &=& 2 {\cal T}_{35c} + {\cal
T}_{6,PI}^{(1)},\nonumber\\ {\cal T}_{\Xi_{cc}^{+}} &=& 2 {\cal
T}_{35c} + {\cal T}_{6,WS}^{(2)},\nonumber\\ {\cal
T}_{\Omega_{cc}^{+}} &=& 2 {\cal T}_{35c} + {\cal
T}_{6,PI}^{(3)},\nonumber\\ {\cal T}_{\Xi_{bc}^{+}} &=& {\cal
T}_{35b} + {\cal T}_{35c} + {\cal T}_{6,PI}^{(4)} + {\cal
T}_{6,WS}^{(4)},\nonumber\\ {\cal T}_{\Xi_{bc}^{0}} &=& {\cal
T}_{35b} + {\cal T}_{35c} + {\cal T}_{6,PI}^{(5)} + {\cal
T}_{6,WS}^{(5)},\nonumber \\ {\cal T}_{\Omega_{bc}^{0}} &=& {\cal
T}_{35b} + {\cal T}_{35c} + {\cal T}_{6,PI}^{(6)} + {\cal
T}_{6,WS}^{(6)},\nonumber \\ {\cal T}_{\Xi_{bb}^{0}} &=& 2 {\cal
T}_{35b} + {\cal T}_{6,WS}^{(7)},\nonumber\\ {\cal
T}_{\Xi_{bb}^{-}} &=& 2 {\cal T}_{35b} + {\cal
T}_{6,PI}^{(8)},\nonumber \\ {\cal T}_{\Omega_{bb}^{-}} &=& 2
{\cal T}_{35b} + {\cal T}_{6,PI}^{(9)},\nonumber
\end{eqnarray}
where the $35$-labelled terms account for the operators of dimension three
$O_{3Q^i}$ and five $O_{GQ^i}$, the $6$-marked terms correspond to the effects
of Pauli interference and weak scattering. The explicit formulae for these
contributions have the following form:
\begin{equation}
{\cal T}_{35b} = \Gamma_{b,spec}\bar bb - \frac{\Gamma_{0b}}{m_b^2}[2P_{c1} +
P_{c\tau 1} + K_{0b}(P_{c1} + P_{cc1}) + K_{2b}(P_{c2} + P_{cc2})]O_{Gb},
\label{5}
\end{equation}
\begin{equation}
{\cal T}_{35c} = \Gamma_{c,spec}\bar cc - \frac{\Gamma_{0c}}{m_c^2}[(2 +
K_{0c})P_{s1} + K_{2c}P_{s2}]O_{Gc}, \label{6}
\end{equation}
where
\begin{equation}
\Gamma_{0b} = \frac{G_F^2m_b^5}{192{\pi}^3}|V_{cb}|^2\qquad , \Gamma_{0c} =
\frac{G_F^2m_c^5}{192{\pi}^3}
\end{equation}
with $K_{0Q} = C_{-}^2 + 2C_{+}^2,~K_{2Q} = 2(C_{+}^2 - C_{-}^2)$, and
$\Gamma_{Q,spec}$ denotes the spectator width (see \cite{bigi,9,10,11}):
\begin{equation}
P_{c1} = (1-y)^4,\quad P_{c2} = (1-y)^3,
\end{equation}
\begin{eqnarray}
P_{c\tau 1} &=& \sqrt{1-2(r+y)+(r-y)^2}[1 - 3(r+y) + 3(r^2+y^2) - r^3 - y^3
-4ry + \nonumber\\
&& 7ry(r+y)] + 12r^2y^2\ln\frac{(1-r-y+\sqrt{1-2(r+y)+(r-y)^2})^2}{4ry}
\end{eqnarray}
\begin{equation}
P_{cc1} = \sqrt{1-4y}(1 - 6y + 2y^2 + 12y^3)
24y^4\ln\frac{1+\sqrt{1-4y}}{1-\sqrt{1-4y}}
\end{equation}
\begin{equation}
P_{cc2} = \sqrt{1-4y}(1 + \frac{y}{2} + 3y^2)
- 3y(1-2y^2)\ln\frac{1+\sqrt{1-4y}}{1-\sqrt{1-4y}}
\end{equation}
\noindent
where $y = \frac{m_c^2}{m_b^2}$ and $r = m_{\tau}^2/m_b^2$. The functions
$P_{s1} (P_{s2})$ can be obtained from $P_{c1} (P_{c2})$ by the
substitution $y = m_s^2/m_c^2$. In the $b$-quark decays, we neglect the
value $m_s^2/m_b^2$ and suppose $m_s = 0$.

The calculation of both the Pauli interference effect for the products of heavy
quark decays with the quarks in the initial state and the weak scattering of
quarks, composing the hadron, depends on the quark contents of baryons and
results in:
\begin{eqnarray}
{\cal T}_{6,PI}^{(1)} &=& 2 {\cal T}_{PI,u\bar d}^c \\
{\cal T}_{6,WS}^{(2)} &=& 2 {\cal T}_{WS,cd} \\
{\cal T}_{6,PI}^{(3)} &=& 2 {\cal T}_{PI,u\bar d}^{c'} + 2 \sum_{l} {\cal
T}_{PI,\nu_l \bar l}^c  \\
{\cal T}_{6,PI}^{(4)} &=& {\cal T}_{PI,u\bar d}^c + {\cal T}_{PI,s\bar c}^b +
{\cal T}_{PI,d\bar u}^b + \sum_l{\cal T}_{PI,l\bar\nu_l}^b\\
{\cal T}_{6,WS}^{(4)} &=& {\cal T}_{WS,bu} + {\cal T}_{WS,bc}\\
{\cal T}_{6,PI}^{(5)} &=& {\cal T}_{PI,s\bar c}^b + {\cal T}_{PI,d\bar u}^b +
{\cal T}_{PI,d\bar u}^{'b} + \sum_l{\cal T}_{PI,l\bar\nu_l}^b\\
{\cal T}_{6,WS}^{(5)} &=& {\cal T}_{WS,cd} + {\cal T}_{WS,bc} \\
{\cal T}_{6,PI}^{(6)} &=& {\cal T}_{PI,u\bar d}^{c'} + \sum_{l} {\cal
T}_{PI,\nu_l \bar l}^c + {\cal T}_{PI,s\bar c}^b + {\cal T}_{PI,d\bar u}^b +
\sum_l{\cal T}_{PI,l\bar\nu_l}^b + {\cal T}_{PI,s\bar c}^{'b}\\
{\cal T}_{6,WS}^{(6)} &=& {\cal T}_{WS,bc} + {\cal T}_{WS,cs}\\
{\cal T}_{6,WS}^{(7)} &=& 2 {\cal T}_{WS,bu} \\
{\cal T}_{6,PI}^{(8)} &=& 2 {\cal T}_{PI,d\bar u}^{'b} \\
{\cal T}_{6,PI}^{(9)} &=& 2 {\cal T}_{PI,s\bar c}^{'b}
\end{eqnarray}
\noindent
so that
\begin{eqnarray}
{\cal T}_{PI,s\bar c}^b &=&
-\frac{G_F^2|V_{cb}|^2}{4\pi}m_b^2(1-\frac{m_c}{m_b})^2\nonumber\\
&& ([(\frac{(1-z_{-})^2}{2}- \frac{(1-z_{-})^3}{4})
(\bar b_i\gamma_{\alpha}(1-\gamma_5)b_i)(\bar
c_j\gamma^{\alpha}(1-\gamma_5)c_j) + \nonumber\\
&& (\frac{(1-z_{-})^2}{2} -
\frac{(1-z_{-})^3}{3})(\bar b_i\gamma_{\alpha}\gamma_5 b_i)(\bar
c_j\gamma^{\alpha}(1-\gamma_5)c_j)] \label{16}
\\&& [(C_{+} - C_{-})^2 +
\frac{1}{3}(1-k^{\frac{1}{2}})(5C_{+}^2+C_{-}^2+6C_{-}C_{+})]+ \nonumber\\
&& [(\frac{(1-z_{-})^2}{2} - \frac{(1-z_{-})^3}{4})(\bar
b_i\gamma_{\alpha}(1-\gamma_5)b_j)(\bar c_j\gamma^{\alpha}(1-\gamma_5)c_i) +
\nonumber\\
&&  (\frac{(1-z_{-})^2}{2} - \frac{(1-z_{-})^3}{3})(\bar
b_i\gamma_{\alpha}\gamma_5b_j)(\bar
c_j\gamma^{\alpha}(1-\gamma_5)c_i)]k^{\frac{1}{2}}(5C_{+}^2+C_{-}^2+6C_{-}C_{+}
)),\nonumber\\
{\cal T}_{PI,\tau\bar\nu_{\tau}}^b &=&
-\frac{G_F^2|V_{cb}|^2}{\pi}m_b^2(1-\frac{m_c}{m_b})^2\nonumber\\
&& [(\frac{(1-z_{\tau})^2}{2} - \frac{(1-z_{\tau})^3}{4})(\bar
b_i\gamma_{\alpha}(1-\gamma_5)b_j)(\bar c_j\gamma^{\alpha}(1-\gamma_5)c_i) +
\label{17}\\
&&  (\frac{(1-z_{\tau})^2}{2} - \frac{(1-z_{\tau})^3}{3})(\bar
b_i\gamma_{\alpha}\gamma_5b_j)(\bar
c_j\gamma^{\alpha}(1-\gamma_5)c_i)],\nonumber\\
{\cal T}_{PI,d\bar u}^{b'} &=&
-\frac{G_F^2|V_{cb}|^2}{4\pi}m_b^2(1-\frac{m_d}{m_b})^2\nonumber\\
&& ([(\frac{(1-z_{-})^2}{2}- \frac{(1-z_{-})^3}{4})
(\bar b_i\gamma_{\alpha}(1-\gamma_5)b_i)(\bar
d_j\gamma^{\alpha}(1-\gamma_5)d_j) + \nonumber\\
&& (\frac{(1-z_{-})^2}{2} -
\frac{(1-z_{-})^3}{3})(\bar b_i\gamma_{\alpha}\gamma_5 b_i)(\bar
d_j\gamma^{\alpha}(1-\gamma_5)d_j)] \label{18}
\\&& [(C_{+} + C_{-})^2 +
\frac{1}{3}(1-k^{\frac{1}{2}})(5C_{+}^2+C_{-}^2-6C_{-}C_{+})]+ \nonumber\\
&& [(\frac{(1-z_{-})^2}{2} - \frac{(1-z_{-})^3}{4})(\bar
b_i\gamma_{\alpha}(1-\gamma_5)b_j)(\bar d_j\gamma^{\alpha}(1-\gamma_5)d_i) +
\nonumber\\
&&  (\frac{(1-z_{-})^2}{2} - \frac{(1-z_{-})^3}{3})(\bar
b_i\gamma_{\alpha}\gamma_5b_j)(\bar
d_j\gamma^{\alpha}(1-\gamma_5)d_i)]k^{\frac{1}{2}}(5C_{+}^2+C_{-}^2-6C_{-}C_{+}
)),\nonumber\\
{\cal T}_{PI,s\bar c}^{b'} &=&
-\frac{G_F^2|V_{cb}|^2}{16\pi}m_b^2(1-\frac{m_s}{m_b})^2\sqrt{(1-4z_{-})}
\nonumber\\
&& ([(1-z_{-})(\bar b_i\gamma_{\alpha}(1-\gamma_5)b_i)(\bar
s_j\gamma^{\alpha}(1-\gamma_5)s_j) + \nonumber\\
&&\frac{2}{3}(1+2z_{-})(\bar b_i\gamma_{\alpha}\gamma_5 b_i)(\bar
s_j\gamma^{\alpha}(1-\gamma_5)s_j)] \label{181}
\\&& [(C_{+} + C_{-})^2 +
\frac{1}{3}(1-k^{\frac{1}{2}})(5C_{+}^2+C_{-}^2-6C_{-}C_{+})]+ \nonumber\\
&& [(1-z_{-})(\bar
b_i\gamma_{\alpha}(1-\gamma_5)b_j)(\bar s_j\gamma^{\alpha}(1-\gamma_5)s_i) +
\nonumber\\
&&\frac{2}{3}(1+2z_{-})(\bar
b_i\gamma_{\alpha}\gamma_5b_j)(\bar
s_j\gamma^{\alpha}(1-\gamma_5)s_i)]k^{\frac{1}{2}}(5C_{+}^2+C_{-}^2-6C_{-}C_{+}
)),\nonumber\\
{\cal T}_{PI,u\bar d}^c &=&
-\frac{G_F^2}{4\pi}m_c^2(1-\frac{m_u}{m_c})^2\nonumber\\
&& ([(\frac{(1-z_{-})^2}{2}- \frac{(1-z_{-})^3}{4})
(\bar c_i\gamma_{\alpha}(1-\gamma_5)c_i)(\bar
u_j\gamma^{\alpha}(1-\gamma_5)u_j) + \nonumber\\
&& (\frac{(1-z_{-})^2}{2} -
\frac{(1-z_{-})^3}{3})(\bar c_i\gamma_{\alpha}\gamma_5 c_i)(\bar
u_j\gamma^{\alpha}(1-\gamma_5)u_j)] \label{19}
\\&& [(C_{+} + C_{-})^2 +
\frac{1}{3}(1-k^{\frac{1}{2}})(5C_{+}^2+C_{-}^2-6C_{-}C_{+})]+ \nonumber\\
&& [(\frac{(1-z_{-})^2}{2} - \frac{(1-z_{-})^3}{4})(\bar
c_i\gamma_{\alpha}(1-\gamma_5)c_j)(\bar u_j\gamma^{\alpha}(1-\gamma_5)u_i) +
\nonumber\\
&&  (\frac{(1-z_{-})^2}{2} - \frac{(1-z_{-})^3}{3})(\bar
c_i\gamma_{\alpha}\gamma_5c_j)(\bar
u_j\gamma^{\alpha}(1-\gamma_5)u_i)]k^{\frac{1}{2}}(5C_{+}^2+C_{-}^2-6C_{-}C_{+}
)),\nonumber\\
{\cal T}_{PI,u\bar d}^{c'} &=&
-\frac{G_F^2}{4\pi}m_c^2(1-\frac{m_s}{m_c})^2\nonumber\\
&& ([\frac{1}{4}(\bar c_i\gamma_{\alpha}(1-\gamma_5)c_i)(\bar
s_j\gamma^{\alpha}(1-\gamma_5)s_j) +
\frac{1}{6}(\bar c_i\gamma_{\alpha}\gamma_5 c_i)(\bar
s_j\gamma^{\alpha}(1-\gamma_5)s_j)] \label{191}
\\&& [(C_{+} - C_{-})^2 +
\frac{1}{3}(1-k^{\frac{1}{2}})(5C_{+}^2+C_{-}^2+6C_{-}C_{+})]+ \nonumber\\
&& [\frac{1}{4}(\bar
c_i\gamma_{\alpha}(1-\gamma_5)c_j)(\bar s_j\gamma^{\alpha}(1-\gamma_5)s_i) +
\nonumber\\
&&\frac{1}{6}(\bar
c_i\gamma_{\alpha}\gamma_5c_j)(\bar
s_j\gamma^{\alpha}(1-\gamma_5)s_i)]k^{\frac{1}{2}}(5C_{+}^2+C_{-}^2+6C_{-}C_{+}
)),\nonumber\\
{\cal T}_{PI,\nu_{\tau}\bar\tau}^c &=&
-\frac{G_F^2}{\pi}m_c^2(1-\frac{m_s}{m_c})^2\nonumber\\
&& [(\frac{(1-z_{\tau})^2}{2} - \frac{(1-z_{\tau})^3}{4})(\bar
c_i\gamma_{\alpha}(1-\gamma_5)c_j)(\bar s_j\gamma^{\alpha}(1-\gamma_5)s_i) +
\label{192}\\
&&  (\frac{(1-z_{\tau})^2}{2} - \frac{(1-z_{\tau})^3}{3})(\bar
c_i\gamma_{\alpha}\gamma_5c_j)(\bar
s_j\gamma^{\alpha}(1-\gamma_5)s_i)],\nonumber\\
{\cal T}_{WS,bc} &=&
\frac{G_F^2|V_{cb}|^2}{4\pi}m_b^2(1+\frac{m_c}{m_b})^2(1-z_{+})^2[(C_{+}^2 +
C_{-}^2 +
 \frac{1}{3}(1 - k^{\frac{1}{2}})(C_{+}^2 - C_{-}^2))\nonumber\\
&&(\bar b_i\gamma_{\alpha}(1
- \gamma_5)b_i)(\bar c_j\gamma^{\alpha}(1 - \gamma_5)c_j) + \label{20}\\
&& k^{\frac{1}{2}}(C_{+}^2 - C_{-}^2)(\bar b_i\gamma_{\alpha}(1 - \gamma_5)b_j)
(\bar c_j\gamma^{\alpha}(1 - \gamma_5)c_i)],\nonumber\\
{\cal T}_{WS,bu} &=&
\frac{G_F^2|V_{cb}|^2}{4\pi}m_b^2(1+\frac{m_u}{m_b})^2(1-z_{+})^2[(C_{+}^2 +
C_{-}^2 +
 \frac{1}{3}(1 - k^{\frac{1}{2}})(C_{+}^2 - C_{-}^2))\nonumber\\
&&(\bar b_i\gamma_{\alpha}(1
- \gamma_5)b_i)(\bar u_j\gamma^{\alpha}(1 - \gamma_5)u_j) + \label{21}\\
&& k^{\frac{1}{2}}(C_{+}^2 - C_{-}^2)(\bar b_i\gamma_{\alpha}(1 - \gamma_5)b_j)
(\bar u_j\gamma^{\alpha}(1 - \gamma_5)u_i)],\nonumber\\
{\cal T}_{WS,cd} &=&
\frac{G_F^2}{4\pi}m_c^2(1+\frac{m_d}{m_c})^2(1-z_{+})^2[(C_{+}^2 + C_{-}^2 +
 \frac{1}{3}(1 - k^{\frac{1}{2}})(C_{+}^2 - C_{-}^2))\nonumber\\
&&(\bar c_i\gamma_{\alpha}(1
- \gamma_5)c_i)(\bar d_j\gamma^{\alpha}(1 - \gamma_5)d_j) + \label{22}\\
&& k^{\frac{1}{2}}(C_{+}^2 - C_{-}^2)(\bar c_i\gamma_{\alpha}(1 - \gamma_5)c_j)
(\bar d_j\gamma^{\alpha}(1 - \gamma_5)d_i)],\nonumber
\end{eqnarray}
\begin{eqnarray}
{\cal T}_{PI,d\bar u}^b &=& {\cal T}_{PI,s\bar c}^b~(z_{-}\to 0)\\
{\cal T}_{PI,e\bar\nu_e}^b &=& {\cal T}_{PI,\mu\bar\nu_{\mu}}^b =
{\cal T}_{PI,\tau\bar\nu_{\tau}}^b~(z_{\tau}\to 0),\\
{\cal T}_{PI,\nu_e\bar e}^c &=& {\cal T}_{PI,\nu_{\mu}\bar\mu}^c =
{\cal T}_{PI,\nu_{\tau}\bar\tau}^c~(z_{\tau}\to 0),
\end{eqnarray}
where the following notation has been used:
\begin{eqnarray}
\mbox{in Eq.}~(\ref{16})&& z_{-} = \frac{m_c^2}{(m_b-m_c)^2},\quad
k = \frac{\alpha_s(\mu)}{\alpha_s(m_b-m_c)},\nonumber\\ \mbox{in
Eq.}~(\ref{17})&& z_{\tau} =
\frac{m_{\tau}^2}{(m_b-m_c)^2},\nonumber\\ \mbox{in
Eq.}~(\ref{18})&& z_{-} = \frac{m_c^2}{(m_b-m_d)^2},\quad k =
\frac{\alpha_s(\mu)}{\alpha_s(m_b-m_d)},\nonumber\\ \mbox{in
Eq.}~(\ref{181})&& z_{-} = \frac{m_c^2}{(m_b-m_s)^2},\quad k =
\frac{\alpha_s(\mu)}{\alpha_s(m_b-m_s)},\nonumber\\ \mbox{in
Eq.}~(\ref{19})&& z_{-} = \frac{m_s^2}{(m_c-m_u)^2},\quad k =
\frac{\alpha_s(\mu)}{\alpha_s(m_c-m_u)},\nonumber\\ \mbox{in
Eq.}~(\ref{191})&&  k =
\frac{\alpha_s(\mu)}{\alpha_s(m_c-m_s)},\nonumber\\ \mbox{in
Eq.}~(\ref{192})&& z_{\tau} =
\frac{m_{\tau}^2}{(m_c-m_s)^2},\nonumber\\ \mbox{in
Eq.}~(\ref{20})&& z_{+} = \frac{m_c^2}{(m_b+m_c)^2},\quad k =
\frac{\alpha_s(\mu)}{\alpha_s(m_b+m_c)},\nonumber\\ \mbox{in
Eq.}~(\ref{21})&& z_{+} = \frac{m_c^2}{(m_b+m_u)^2},\quad k =
\frac{\alpha_s(\mu)}{\alpha_s(m_b+m_u)},\nonumber\\ \mbox{in
Eq.}~(\ref{22})&& z_{+} = \frac{m_s^2}{(m_c+m_d)^2},\quad k =
\frac{\alpha_s(\mu)}{\alpha_s(m_c+m_d)}.\nonumber
\end{eqnarray}
In the evolution of coefficients $C_{+}$ and
$C_{-}$, we have taken into account the threshold effects, connected to the
heavy quark masses.

In expressions (\ref{5}) and  (\ref{6}), the scale $\mu$ has been
taken approximately equal to $m_c$. In the Pauli interference
term, we suggest that the scale can be determined on the basis of
the agreement of the experimentally known difference between the
lifetimes of $\Lambda_c$, $\Xi_c^{+}$ and $\Xi_c^{0}$ with the
theoretical predictions in the framework described
above\footnote{A more extended description is presented in
\cite{DHD1}.}. In any case, the choice of the normalization scale
leads to uncertainties in the final results. Theoretical accuracy
can be improved by the calculation of next-order corrections in
the powers of QCD coupling constant.

The coefficients of leading terms, represented by operators $\bar bb$ and $\bar
cc$, are equivalent to the widths fot the decays of free quarks and are known
in the next-to-leading logarithmic approximation of QCD \cite{12,13,14,15,16},
including the mass corrections in the final state with the charmed quark and
$\tau$-lepton \cite{16} in the decays of $b$-quark and with the strange quark
mass for the decays of $c$-quark. In the numerical estimates, we include these
corrections and mass effects, but we neglect the decay modes suppressed by the
Cabibbo angle, and also the strange quark mass effects in $b$ decays.

The expressions for the contribution of operator $\sum_{i=1}^2O_{GQ^i}$ are
known in the leading logarithmic approximation. The expressions for the
contributions of operators with the dimension 6 have been calculated by us with
account of masses in the final states together with the logarithmic
renormalization of the effective lagrangian for the non-relativistic heavy
quarks at energies less than the heavy quark masses.

Thus, the calculation of lifetimes for the baryons $\Xi_{QQ'}^{\diamond}$ is
reduced to the problem of evaluating the matrix elements of operators, which is
the subject of next section.

\section{Matrix elements in NRQCD approximation.}

By using the equations of motion, the matrix element of operator $\bar Q^jQ^j$
can be expanded in series over the powers of ${1}/{m_{Q^j}}$:
\begin{eqnarray}
\langle \Xi_{QQ'}^{\diamond}|\bar
Q^jQ^j|\Xi_{QQ'}^{\diamond}\rangle _{norm} = 1 - \frac{\langle
\Xi_{QQ'}^{\diamond}|\bar
Q^j[(i\boldsymbol{D})^2-(\frac{i}{2}\sigma
G)]Q^j|\Xi_{QQ'}^{\diamond}\rangle_{norm}}{2m_{Q^j}^2} +
O(\frac{1}{m_{Q^j}^3}).
\end{eqnarray}
Thus, we have to estimate the matrix elements of operators from
the following list only:
\begin{eqnarray}
&& \bar Q^j(i\boldsymbol{ D})^2Q^j,\quad (\frac{i}{2})\bar
Q^j\sigma GQ^j,\quad \bar Q^j\gamma_{\alpha}(1-\gamma_5)Q^j\bar
q\gamma^{\alpha}(1-\gamma_5)q,\nonumber\\ && \bar
Q^j\gamma_{\alpha}\gamma_5Q^j\bar
q\gamma^{\alpha}(1-\gamma_5)q,\quad \bar
Q^j\gamma_{\alpha}\gamma_5Q^j\bar
Q^k\gamma^{\alpha}(1-\gamma_5)Q^k,\\ && \bar
Q^j\gamma_{\alpha}(1-\gamma_5)Q^j\bar
Q^k\gamma^{\alpha}(1-\gamma_5)Q^k.\nonumber
\end{eqnarray}
The meaning of each term in the above list, was already discussed
by us in the previous papers on the decays of doubly heavy baryons
\cite{DHD1,DHD2}, so we omit it here.

Further, employing the NRQCD expansion of operators $\bar Q Q$ and
$\bar Qg\sigma_{\mu\nu}G^{\mu\nu}Q$, we have
\begin{eqnarray}
\bar QQ &=& \Psi_Q^{\dagger}\Psi_Q -
\frac{1}{2m_Q^2}\Psi_Q^{\dagger}(i\boldsymbol{ D})^2\Psi_Q +
\frac{3}{8m_Q^4}\Psi_Q^{\dagger}(i\boldsymbol{ D})^4\Psi_Q
-\nonumber\\ &&
\frac{1}{2m_Q^2}\Psi_Q^{\dagger}g\boldsymbol{\sigma}\boldsymbol{
B}\Psi_Q - \frac{1}{4m_Q^3}\Psi_Q^{\dagger}(\boldsymbol{
D}g\boldsymbol{ E})\Psi_Q + ... \label{32}\\ \bar
Qg\sigma_{\mu\nu}G^{\mu\nu}Q &=&
-2\Psi_Q^{\dagger}g\boldsymbol{\sigma}\boldsymbol{ B}\Psi_Q -
\frac{1}{m_Q}\Psi_Q^{\dagger}(\boldsymbol{ D}g\boldsymbol{
E})\Psi_Q + ... \label{33}
\end{eqnarray}
Here the factorization at scale $\mu$ ($m_{Q} >  \mu > m_{Q}v_{Q}$) is
supposed. We have omitted the term of $\Psi_Q^{\dagger}\boldsymbol{\sigma}
(g\boldsymbol{E} \times \boldsymbol{ D})\Psi_Q$, corresponding to the
spin-orbital interactions, which are not essential for the basic state of
baryons under consideration. The field $\Psi_Q$ has the standard
non-relativistic normalization.

Now we would like to make some comments on the difference between
the descriptions of interactions of the heavy quark with the
light and heavy heavy ones. As well known, in the doubly heavy subsystem there
is an additional parameter which is the relative velocity of quarks. It
introduces the energy scale equal to $m_Q v$. Therefore, the Darwin term
($\boldsymbol{ D}\boldsymbol{ E}$) in the heavy subsystem stands
in the same order of inverse heavy quark mass in comparison with
the chromomagnetic term ($\boldsymbol{\sigma}\boldsymbol{ B}$)
(they have the same power in the velocity $v$). This statement
becomes evident if we apply the scaling rules of NRQCD \cite{4}:
$$ \Psi_Q\sim (m_Qv_Q)^{\frac{3}{2}},\quad|\boldsymbol{ D}|\sim
m_Qv_Q,\quad gE\sim m_Q^2v_Q^3,\quad gB\sim m_Q^2v_Q^4,\quad g\sim
v_Q^{\frac{1}{2}}. $$ For the interaction of heavy quark with the
light one, there is no such small velocity parameter, so that the
Darwin term is suppressed by the additional factor of $k/m_Q\sim
\Lambda_{QCD}/m_Q$.

Further, the phenomenological experience with the potential quark
models shows, that the kinetic energy of quarks practically does
not depend on the quark contents of system, and it is determined
by the color structure of state. So, we suppose that the kinetic
energy is equal to $T = m_dv_d^2/2 + m_lv_l^2/2$ for the
quark-diquark system, and it is $T/2 = m_{b}v_{b}^2/2 +
m_{c}v_{c}^2/2$ in the diquark (the color factor of 1/2). Then
\begin{equation}
\frac{\langle
\Xi_{QQ'}^{\diamond}|\Psi_Q^{\dagger}(i\boldsymbol{D})^2\Psi_Q|\Xi_{QQ'}^{
\diamond} \rangle }{2M_{\Xi_{QQ'}^{\diamond}}m_Q^2}\simeq
v_Q^2\simeq
\frac{2m_qT}{(m_q+m_{Q'}+m_{Q})(m_{Q'}+m_{Q})}+\frac{m_{Q'}T}{m_Q(m_Q+m_{Q'
})},
\end{equation}
\begin{equation}
\frac{\langle
\Xi_{QQ'}^{\diamond}|\Psi_{Q'}^{\dagger}(i\boldsymbol{D})^2\Psi_{Q'}|\Xi_{QQ'}^
{
\diamond} \rangle }{2M_{\Xi_{QQ'}^{\diamond}}m_{Q'}^2}\simeq
v_{Q'}^2\simeq
\frac{2m_qT}{(m_q+m_{Q'}+m_{Q})(m_{Q'}+m_{Q})}+\frac{m_QT}{m_{Q'}(m_Q+m_{Q'})},
\end{equation}
where the diquark terms dominate certainly. Applying the quark-diquark
approximation and relating the matrix element of chromomagnetic interaction of
diquark with the light quark to the mass difference between the exited and
ground states $M_{\Xi_{QQ'}^{\diamond *}} - M_{\Xi_{QQ'}^{\diamond}}$, we have
\begin{eqnarray}
\frac{\langle \Xi_{cc}^{\diamond}|\bar cc|\Xi_{cc}^{\diamond}\rangle
}{2M_{\Xi_{cc}^{\diamond}}} &=& 1 -
\frac{1}{2}v_c^2 -
\frac{1}{3}\frac{M_{\Xi_{cc}^{\diamond *}}-M_{\Xi_{cc}^{\diamond}}}{m_c}
- \frac{5 g^2}{18 m_c^3}|\Psi (0)|^2 + ... \nonumber\\
&\approx& 1 - 0.073 -0.025 - 0.009 +\ldots   \\
\frac{\langle \Omega_{cc}^{\diamond}|\bar cc|\Omega_{cc}^{\diamond}\rangle
}{2M_{\Omega_{cc}^{\diamond}}} &=& 1 -
\frac{1}{2}v_c^2 -
\frac{1}{3}\frac{M_{\Omega_{cc}^{\diamond *}}-M_{\Omega_{cc}^{\diamond}}}{m_c}
- \frac{5 g^2}{18 m_c^3}|\Psi (0)|^2 + ... \nonumber\\
&\approx& 1 - 0.078 -0.025 - 0.009 +\ldots \\
\frac{\langle \Xi_{bc}^{\diamond}|\bar
cc|\Xi_{bc}^{\diamond}\rangle }{2M_{\Xi_{bc}^{\diamond}}} &=& 1 -
\frac{1}{2}v_c^2 + \frac{g^2}{3m_bm_c^2}|\Psi^d (0)|^2 -
\frac{1}{6m_c^3}g^2|\Psi^d (0)|^2 + \ldots
\nonumber\\ &\approx& 1 - 0.098 + 0.006 - 0.010\ldots \\
\frac{\langle \Omega_{bc}^{\diamond}|\bar
cc|\Omega_{bc}^{\diamond}\rangle }{2M_{\Omega_{bc}^{\diamond}}} &=& 1 -
\frac{1}{2}v_c^2 + \frac{g^2}{3m_bm_c^2}|\Psi^d (0)|^2 -
\frac{1}{6m_c^3}g^2|\Psi^d (0)|^2 + \ldots
\nonumber\\ &\approx& 1 - 0.099 + 0.006 - 0.010\ldots  \\
\frac{\langle \Xi_{bb}^{\diamond}|\bar bb|\Xi_{bb}^{\diamond}\rangle
}{2M_{\Xi_{bb}^{\diamond}}} &=& 1 -
\frac{1}{2}v_b^2 -
\frac{1}{3}\frac{M_{\Xi_{bb}^{\diamond *}}-M_{\Xi_{bb}^{\diamond}}}{m_b}
- \frac{5 g^2}{18 m_b^3}|\Psi (0)|^2 + ... \nonumber\\
&\approx& 1 - 0.021 -0.003 - 0.002 +\ldots   \\
\frac{\langle \Omega_{bb}^{\diamond}|\bar bb|\Omega_{bb}^{\diamond}\rangle
}{2M_{\Omega_{bb}^{\diamond}}} &=& 1 -
\frac{1}{2}v_b^2 -
\frac{1}{3}\frac{M_{\Omega_{bb}^{\diamond *}}-M_{\Omega_{bb}^{\diamond}}}{m_b}
- \frac{5 g^2}{18 m_b^3}|\Psi (0)|^2 + ... \nonumber\\
&\approx& 1 - 0.021 -0.003 - 0.002 +\ldots
\end{eqnarray}
The numerical values of parameters used in the calculations above are given in
the next section. Our presentation here is less detailed than in previous
papers \cite{DHD1,DHD2}. However, we hope, that the interested reader can find
there all needed details.

Analogous expressions may be obtained for the matrix elements of
operator $Qg\sigma_{\mu\nu}G^{\mu\nu}Q$
\begin{eqnarray}
\frac{\langle \Xi_{cc}^{\diamond}|\bar
cg\sigma_{\mu\nu}G^{\mu\nu}c|\Xi_{cc}^{\diamond}\rangle
}{2M_{\Xi_{cc}^{\diamond}}m_c^2} &=&
-\frac{4}{3}\frac{(M_{\Xi_{cc}^{\diamond *}} -
M_{\Xi_{cc}^{\diamond}})}{m_c}
- \frac{7g^2}{9m_c^3}|\Psi^d (0)|^2 \approx -0.124, \\
\frac{\langle \Xi_{bc}^{\diamond}|\bar
cg\sigma_{\mu\nu}G^{\mu\nu}c|\Xi_{bc}^{\diamond}\rangle
}{2M_{\Xi_{bc}^{\diamond}}m_c^2} &=& \frac{4g^2}{3m_b
m_c^2}|\Psi^d (0)|^2 - \frac{g^2}{3m_c^3}|\Psi^d (0)|^2 \approx
0.005, \\
\frac{\langle \Xi_{bb}^{\diamond}|\bar
bg\sigma_{\mu\nu}G^{\mu\nu}b|\Xi_{bb}^{(\diamond)}\rangle
}{2M_{\Xi_{bb}^{\diamond}}m_b^2} &=&
-\frac{4}{3}\frac{(M_{\Xi_{bb}^{\diamond *}} -
M_{\Xi_{bb}^{\diamond}})}{m_b} - \frac{7g^2}{9m_b^3}|\Psi^d (0)|^2 \approx
-0.189, \\
\langle \Omega_{QQ'}|\bar
cg\sigma_{\mu\nu}G^{\mu\nu}c|\Omega_{QQ'}\rangle &=&
\langle \Xi_{QQ'}^{\diamond}|\bar
cg\sigma_{\mu\nu}G^{\mu\nu}c|\Xi_{QQ'}^{\diamond}\rangle
\end{eqnarray}
The permutations of quark masses lead to the required expressions
for the operators of $\bar bb$ and $\bar
bg\sigma_{\mu\nu}G^{\mu\nu}b$.

For the four quark operators, determining the Pauli interference
and the weak scattering, we use the estimates in the framework of
non-relativistic potential model \cite{DHD1,DHD2}:
\begin{eqnarray}
\langle \Xi_{cc}^{\diamond}|(\bar c\gamma_{\mu}(1-\gamma_5)c)(\bar
q\gamma^{\mu}(1-\gamma_5)q)|\Xi_{cc}^{\diamond}\rangle  &=& 12(m_c+m_q)\cdot
|\Psi^{dl}(0)|^2,\\
\langle \Xi_{cc}^{\diamond}|(\bar c\gamma_{\mu}\gamma_5 c)(\bar
q\gamma^{\mu}(1-\gamma_5)q)|\Xi_{cc}^{\diamond}\rangle  &=& 8(m_c+m_q)\cdot
|\Psi^{dl}(0)|^2,\\
\langle \Omega_{cc}|(\bar c\gamma_{\mu}(1-\gamma_5)c)(\bar
s\gamma^{\mu}(1-\gamma_5)s)|\Omega_{cc}\rangle  &=& 12(m_c+m_s)\cdot
|\Psi^{dl}(0)|^2,\\
\langle \Omega_{cc}|(\bar c\gamma_{\mu}\gamma_5 c)(\bar
s\gamma^{\mu}(1-\gamma_5)s)|\Omega_{cc}\rangle  &=& 8(m_c+m_s)\cdot
|\Psi^{dl}(0)|^2,\\
\langle \Xi_{bc}^{\diamond}|(\bar b\gamma_{\mu}(1-\gamma_5)b)(\bar
c\gamma^{\mu}(1-\gamma_5)c)|\Xi_{bc}^{\diamond}\rangle  &=&
8(m_b+m_c)\cdot |\Psi^{d}(0)|^2,\\ \langle
\Xi_{bc}^{\diamond}|(\bar b\gamma_{\mu}\gamma_5b)(\bar
c\gamma^{\mu}(1-\gamma_5)c)|\Xi_{bc}^{\diamond}\rangle  &=&
6(m_b+m_c)\cdot |\Psi^{d}(0)|^2,\\ \langle
\Xi_{bc}^{\diamond}|(\bar b\gamma_{\mu}(1-\gamma_5)b)(\bar
q\gamma^{\mu}(1-\gamma_5)q)|\Xi_{bc}^{\diamond}\rangle  &=&
2(m_b+m_l)\cdot |\Psi^{dl}(0)|^2,\\ \langle
\Xi_{bc}^{\diamond}|(\bar b\gamma_{\mu}\gamma_5b)(\bar
q\gamma^{\mu}(1-\gamma_5)q)|\Xi_{bc}^{\diamond}\rangle  &=& 0,\\
\langle \Xi_{bc}^{\diamond}|(\bar c\gamma_{\mu}(1-\gamma_5)c)(\bar
q\gamma^{\mu}(1-\gamma_5)q)|\Xi_{bc}^{\diamond}\rangle  &=&
2(m_c+m_l)\cdot |\Psi^{dl}(0)|^2,\\ \langle
\Xi_{bc}^{\diamond}|(\bar c\gamma_{\mu}\gamma_5c)(\bar
q\gamma^{\mu}(1-\gamma_5)q)|\Xi_{bc}^{\diamond}\rangle  &=& 0,\\
\langle \Omega_{bc}|(\bar b\gamma_{\mu}(1-\gamma_5)b)(\bar
c\gamma^{\mu}(1-\gamma_5)c)|\Omega_{bc}\rangle  &=&
8(m_b+m_c)\cdot |\Psi^{d}(0)|^2,\\ \langle
\Omega_{bc}|(\bar b\gamma_{\mu}\gamma_5b)(\bar
c\gamma^{\mu}(1-\gamma_5)c)|\Omega_{bc}\rangle  &=&
6(m_b+m_c)\cdot |\Psi^{d}(0)|^2,\\ \langle
\Omega_{bc}|(\bar b\gamma_{\mu}(1-\gamma_5)b)(\bar
s\gamma^{\mu}(1-\gamma_5)s)|\Omega_{bc}\rangle  &=&
2(m_b+m_s)\cdot |\Psi^{dl}(0)|^2,\\ \langle
\Omega_{bc}|(\bar b\gamma_{\mu}\gamma_5b)(\bar
s\gamma^{\mu}(1-\gamma_5)s)|\Omega_{bc}\rangle  &=& 0,\\
\langle \Omega_{bc}|(\bar c\gamma_{\mu}(1-\gamma_5)c)(\bar
s\gamma^{\mu}(1-\gamma_5)s)|\Omega_{bc}\rangle  &=&
2(m_c+m_s)\cdot |\Psi^{dl}(0)|^2,\\ \langle
\Omega_{bc}|(\bar c\gamma_{\mu}\gamma_5c)(\bar
s\gamma^{\mu}(1-\gamma_5)s)|\Omega_{bc}\rangle  &=& 0,\\
\langle \Xi_{bb}^{\diamond}|(\bar b\gamma_{\mu}(1-\gamma_5)b)(\bar
q\gamma^{\mu}(1-\gamma_5)q)|\Xi_{bb}^{\diamond}\rangle  &=& 12(m_b+m_q)\cdot
|\Psi^{dl}(0)|^2,\\
\langle \Xi_{bb}^{\diamond}|(\bar b\gamma_{\mu}\gamma_5 b)(\bar
q\gamma^{\mu}(1-\gamma_5)q)|\Xi_{bb}^{\diamond}\rangle  &=& 8(m_b+m_q)\cdot
|\Psi^{dl}(0)|^2.
\end{eqnarray}
The color structure of wave functions leads to the relations
$$\langle \Xi_{QQ'}^{\diamond}|(\bar Q_iT_{\mu}Q_k)(\bar
q_k\gamma^{\mu}(1-\gamma_5)q_i)|\Xi_{QQ'}^{\diamond}\rangle
=
-\langle \Xi_{QQ'}^{\diamond}|(\bar QT_{\mu}Q)(\bar
q\gamma^{\mu}(1-\gamma_5)q)|\Xi_{QQ'}^{\diamond}\rangle , $$ where
$T_{\mu}$ is an arbitrary spinor matrix.

\section{Numerical estimates}

Performing the numerical calculations of lifetimes for the doubly heavy
baryons, we have used the following set of parameters:
\begin{eqnarray}
&&m_s = 0.2~\mbox{GeV}\quad m_l = 0.~GeV\quad m_s^{*} =
0.45~\mbox{GeV}\quad m_l^{*} = 0.3~\mbox{GeV}\nonumber
\label{parameters}\\ &&\quad\quad |V_{cs}| = 0.9745 \quad |V_{bc}|
= 0.04 \quad T = 0.4~\mbox{GeV}
\\ &&\quad\quad\quad m_c = 1.55~\mbox{GeV}\quad m_b = 5.05~\mbox{GeV}
\end{eqnarray}
The numerical values of diquark wavefunctions at the origin for
baryons under consideration are collected in Table \ref{WF}. The
masses of doubly heavy baryons may be found in Table
\ref{Bmasses}.
\begin{table}[th]
\begin{center}
\begin{tabular}{|c|c|c|c|c|c|c|c|c|c|}
\hline & $\Xi_{cc}^{++}$ & $\Xi_{cc}^{+}$ & $\Omega_{cc}^{+}$ &
$\Xi_{bc}^{+}$ & $\Xi_{bc}^{0}$ & $\Omega_{bc}^{0}$ &
$\Xi_{bb}^{0}$ & $\Xi_{bb}^{-}$ & $\Omega_{bb}^{-}$
\\ \hline
$\Psi^d (0)$, GeV$^{\frac{3}{2}}$ & 0.150 & 0.150 & 0.150 & 0.205
& 0.205 & 0.205 & 0.380 & 0.380 & 0.380 \\ \hline
\end{tabular}
\end{center}
\caption{The values of diquark wavefunctions for the doubly heavy baryons
at the origin.} \label{WF}
\end{table}
\begin{table}[th]
\begin{center}
\begin{tabular}{|c|c|c|c|c|c|c|c|c|c|}
\hline & $\Xi_{cc}^{++}$ & $\Xi_{cc}^{+}$ & $\Omega_{cc}^{+}$ &
$\Xi_{bc}^{+}$ & $\Xi_{bc}^{0}$ & $\Omega_{bc}^{0}$ &
$\Xi_{bb}^{0}$ & $\Xi_{bb}^{-}$ & $\Omega_{bb}^{-}$
\\ \hline
$M$, GeV & 3.478 & 3.478 & 3.578 & 6.82 & 6.82 & 6.92 & 10.093 &
10.093 & 10.193 \\ \hline $M^{*}$, GeV & 3.61 & 3.61 & 3.71 & - &
- & - & 10.193 & 10.193 & 10.293 \\ \hline
\end{tabular}
\end{center}
\caption{The masses of doubly heavy baryons $M$, and $M^{*}$
stands for the mass of the baryon with lowest excited state of
light quark-diquark system.} \label{Bmasses}
\end{table}
The wavefunctions as well as masses for the considered baryons
are taken from \cite{DHS1,DHS2}, where their estimates in the
non-relativistic model with the Buchm\"{u}ller-Tye potential were
done. The $b$-quark mass is obtained from the requirement, that
for any given value of $c$-quark mass the theoretically computed
$B_d$-meson lifetime equals to experimentally measured value.
This matching condition leads to the following approximate
relation
\begin{equation}
m_b = m_c + 3.5~\mbox{GeV}. \label{mb-mc}
\end{equation}
The $c$-quark mass is determined from the analogous matching
procedure for the $B_c$-meson lifetime \cite{Onishchenko}. The
$m_q^{*}$-values in Eq. (\ref{parameters}) represent the
constituent masses for the corresponding light quarks, used by us in
estimations of hadronic matrix elements.\footnote{See \cite{DHD1}
for details.} For the value of light quark-diquark function at the origin we
assume
\begin{equation}
|\Psi^{dl} (0) |^2 = \frac{2}{3}\frac{f_D^2 M_D
k^{-\frac{4}{9}}}{12},
\end{equation}
where $f_D = 170~\mbox{MeV}$. This expression obtained by
performing the steps similar to \cite{Rujula,Cortes} for the
derivation of hyperfine splitting in the light quark-diquark system.
The factor $k^{-\frac{4}{9}}$ accounts for the low energy logarithmic
renormalization of $f_D$ constant. We have used this equation for all
doubly heavy baryons, considered in this paper. Even though we use
this relation to compute central values of lifetimes, the
precise values of wavefunction parameters are under question, so in the
presented results we have allowed for variations.

The renormalization scale $\mu$ is chosen in the following way: $\mu_1 = m_b$
and $\mu_2 = m_c$ in the estimates of Wilson coefficients $C_{+}(\mu)$ and
$C_{-}(\mu)$ for the effective four-fermion weak lagrangian at low energies
with the $b$ and $c$-quarks, correspondingly. For nonspectator effects, which
are the Pauli interference and weak scattering of valence quarks, the
renormalization scale $\mu$ is obtained from the fit of theoretical predictions
for the lifetime differences of baryons $\Lambda_c$, $\Xi_c^{+}$ and
$\Xi_c^{0}$ over the experimental data.

In Table \ref{cc} we present the results of calculations for the
doubly charmed baryons. Together with the total lifetimes of these
baryons we have shown the relative spectator and nonspectator
contributions.
\begin{table}[t]
\begin{center}
\begin{tabular}{|c|c|c|c|}
\hline & $\Xi_{cc}^{++}$ & $\Xi_{cc}^{+}$ & $\Omega_{cc}^{+}$ \\
\hline $\sum c\to s$, ps$^{-1}$ & 3.104 & 3.104 & 3.104 \\ \hline
PI, ps$^{-1}$ & -0.874 & - & 0.621 \\ \hline WS, ps$^{-1}$ & - &
1.776 & - \\ \hline $\tau$, ps & 0.45 & 0.20 & 0.27 \\ \hline
\end{tabular}
\end{center}
\caption{The lifetimes of doubly charmed baryons  together with the
relative spectator and nonspectator contributions to the total
widths.} \label{cc}
\end{table}
From this Table we see the importance of nonspectator effects,
producing huge differences in the values of lifetimes. The
analogous results for other doubly heavy baryons can be found in
Tables \ref{bc} and \ref{bb}.
\begin{table}[b]
\begin{center}
\begin{tabular}{|c|c|c|c|}
\hline & $\Xi_{bc}^{+}$ & $\Xi_{bc}^{0}$ & $\Omega_{bc}^{0}$ \\
\hline $\sum b\to c$, ps$^{-1}$ & 0.632 & 0.632 & 0.631 \\ \hline
$\sum c\to s$, ps$^{-1}$ & 1.511 & 1.511 & 1.509 \\ \hline PI,
ps$^{-1}$ & 0.807 & 0.855 & 0.979 \\ \hline WS, ps$^{-1}$ & 0.653
& 0.795 & 1.713 \\ \hline $\tau$, ps & 0.28 & 0.26 & 0.21 \\
\hline
\end{tabular}
\end{center}
\caption{The lifetimes of $(bcq)$-baryons  together with the relative
spectator and nonspectator contributions to the total widths.}
\label{bc}
\end{table}

\begin{table}[t]
\begin{center}
\begin{tabular}{|c|c|c|c|}
\hline & $\Xi_{bb}^{0}$ & $\Xi_{bb}^{-}$ & $\Omega_{bb}^{-}$ \\
\hline $\sum b\to c$, ps$^{-1}$ & 1.254 & 1.254 & 1.254 \\ \hline
PI, ps$^{-1}$ & - & -0.0130 & -0.0100 \\ \hline WS, ps$^{-1}$ &
0.0189 & - & - \\ \hline $\tau$, ps & 0.79 & 0.80 & 0.80 \\ \hline
\end{tabular}
\end{center}
\caption{The lifetimes of $(bbq)$-baryons  together with the relative
spectator and nonspectator contributions to the total widths.}
\label{bb}
\end{table}
A small comment concerns with the corrections to the spectator decays
of heavy quarks, caused by the motion of heavy quarks inside the
hadron and interactions with the light degrees of freedom. The
corrections due to the quark-gluon operators of dimension 5 are
numerically small \cite{vs}. The most important terms come from the
kinetic energy of heavy quarks.

\begin{center}
\begin{figure}[ph]
\vspace*{-1.cm}
\hbox to 1.5cm
{\hspace*{3.cm}\hfil\mbox{$\tau_{\Xi_{cc}^{++}}$, ps}}
\vspace*{6.5cm} \hbox to 17.5cm {\hfil \mbox{$|\Psi^{dl} (0)|^2$,
GeV$^{3}$}} \vspace*{8.cm}\hspace*{2.5cm} \epsfxsize=12cm
\epsfbox{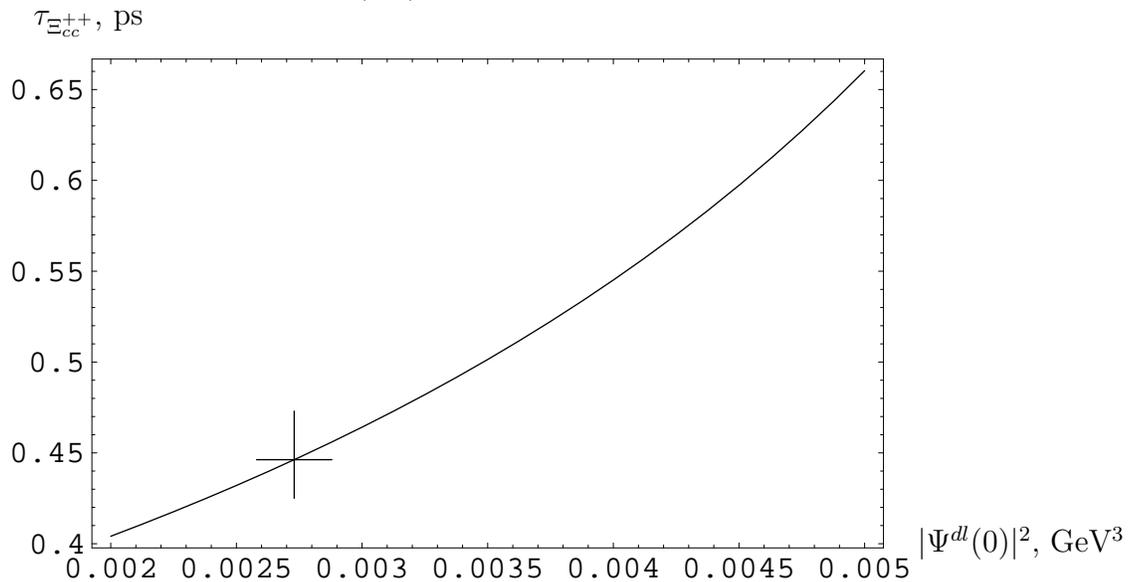} \vspace*{-15.cm}
\caption{The dependence of
$\Xi_{cc}^{++}$-baryon lifetime on the value of wavefunction of
light quark-diquark system at the origin $|\Psi^{dl} (0)|$.}
\label{ccu}
\end{figure}
\end{center}

\begin{center}
\begin{figure}[ph]
\vspace*{-1.cm}
\hbox to 1.5cm
{\hspace*{3.cm}\hfil\mbox{$\tau_{\Xi_{cc}^{+}}$, ps}}
\vspace*{6.5cm} \hbox to 17.5cm {\hfil \mbox{$|\Psi^{dl} (0)|^2$,
GeV$^{3}$}} \vspace*{8.cm}\hspace*{2.5cm} \epsfxsize=12cm
\epsfbox{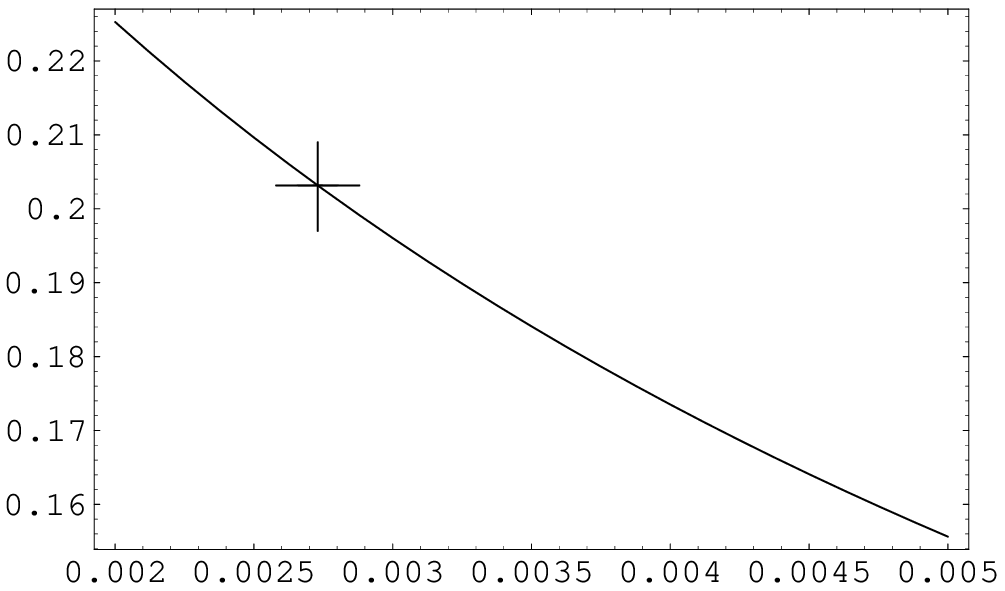} \vspace*{-15.cm}
\caption{The dependence of
$\Xi_{cc}^{+}$-baryon lifetime on the value of wavefunction of
light quark-diquark system at the origin $|\Psi^{dl} (0)|$.}
\label{ccd}
\end{figure}
\end{center}

\begin{center}
\begin{figure}[ph]
\vspace*{-1.cm}
\hbox to 1.5cm
{\hspace*{3.cm}\hfil\mbox{$\tau_{\Omega_{cc}}$, ps}}
\vspace*{6.5cm} \hbox to 17.5cm {\hfil \mbox{$|\Psi^{dl} (0)|^2$,
GeV$^{3}$}} \vspace*{8.cm}\hspace*{2.5cm} \epsfxsize=12cm
\epsfbox{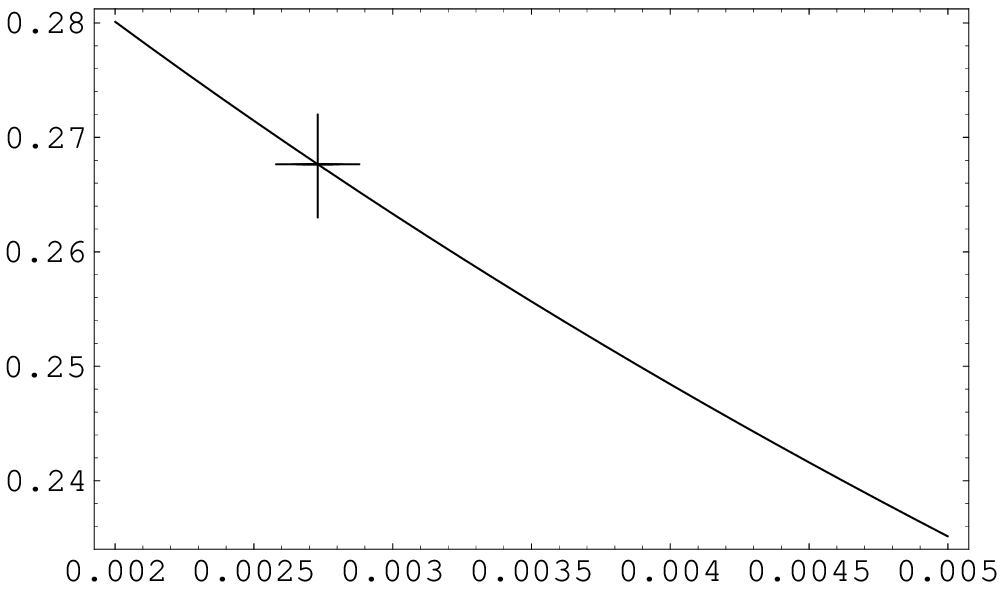} \vspace*{-15.cm}
\caption{The dependence of
$\Omega_{cc}$-baryon lifetime on the value of wavefunction of
light quark-diquark system at the origin $|\Psi^{dl} (0)|$.}
\label{ccs}
\end{figure}
\end{center}

\begin{center}
\begin{figure}[ph]
\vspace*{-1.cm}
\hbox to 1.5cm
{\hspace*{3.cm}\hfil\mbox{$\tau_{\Xi_{bc}^{+}}$, ps}}
\vspace*{6.5cm} \hbox to 17.5cm {\hfil \mbox{$|\Psi^{dl} (0)|^2$,
GeV$^{3}$}} \vspace*{8.cm}\hspace*{2.5cm} \epsfxsize=12cm
\epsfbox{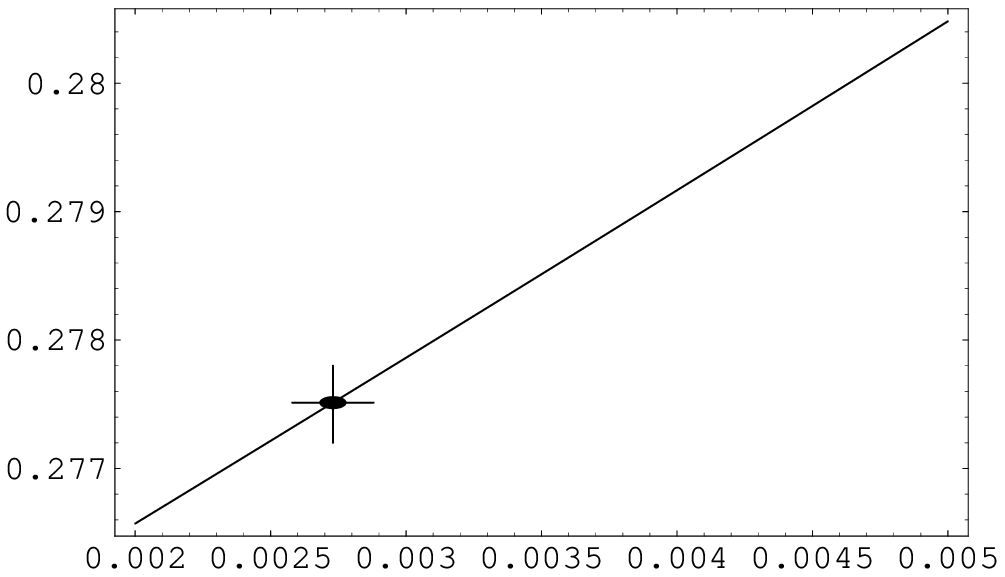} \vspace*{-15.cm}
\caption{The dependence of
$\Xi_{bc}^{+}$-baryon lifetime on the value of wavefunction of
light quark-diquark system at the origin $|\Psi^{dl} (0)|$.}
\label{bcu}
\end{figure}
\end{center}

\begin{center}
\begin{figure}[ph]
\vspace*{-1.cm}
\hbox to 1.5cm
{\hspace*{3.cm}\hfil\mbox{$\tau_{\Xi_{bc}^{0}}$, ps}}
\vspace*{6.5cm} \hbox to 17.5cm {\hfil \mbox{$|\Psi^{dl} (0)|^2$,
GeV$^{3}$}} \vspace*{8.cm}\hspace*{2.5cm} \epsfxsize=12cm
\epsfbox{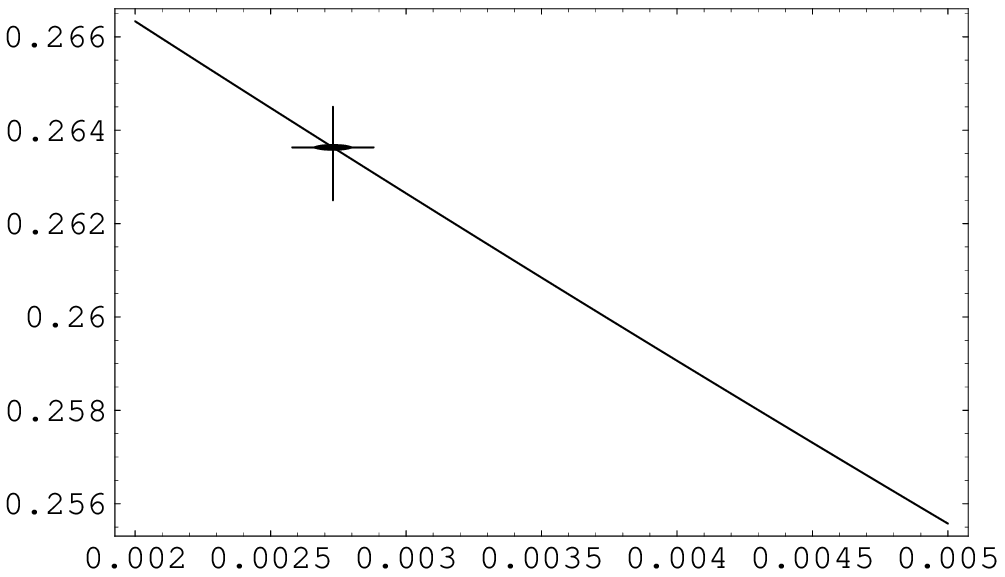} \vspace*{-15.cm}
\caption{The dependence of
$\Xi_{bc}^{0}$-baryon lifetime on the value of wavefunction of
light quark-diquark system at the origin $|\Psi^{dl} (0)|$.}
\label{bcd}
\end{figure}
\end{center}

\begin{center}
\begin{figure}[ph]
\vspace*{-1.cm}
\hbox to 1.5cm
{\hspace*{3.cm}\hfil\mbox{$\tau_{\Omega_{bc}}$, ps}}
\vspace*{6.5cm} \hbox to 17.5cm {\hfil \mbox{$|\Psi^{dl} (0)|^2$,
GeV$^{3}$}} \vspace*{8.cm}\hspace*{2.5cm} \epsfxsize=12cm
\epsfbox{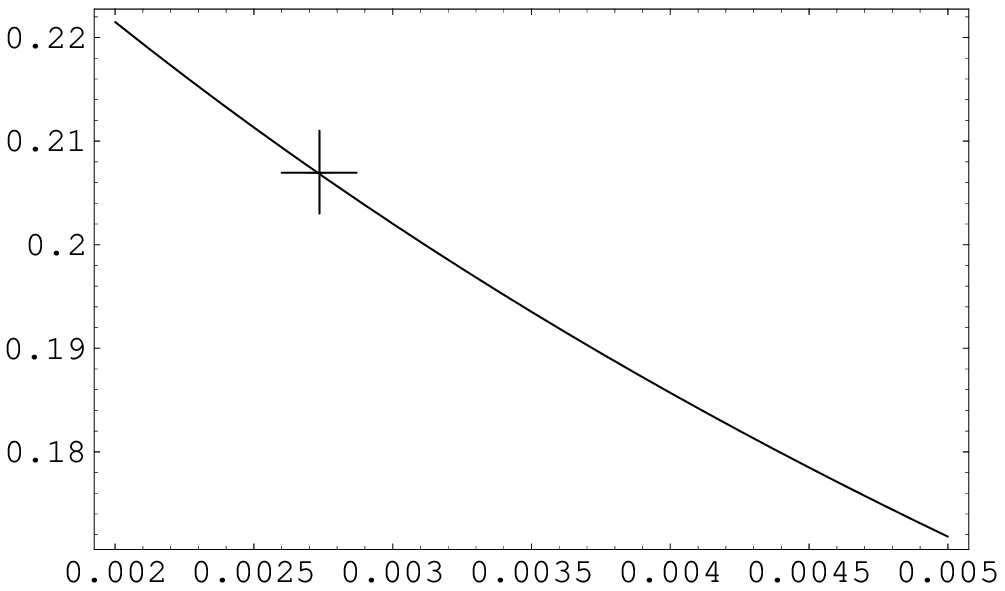} \vspace*{-15.cm}
\caption{The dependence of
$\Omega_{bc}$-baryon lifetime on the value of wavefunction of
light quark-diquark system at theorigin $|\Psi^{dl} (0)|$.}
\label{bcs}
\end{figure}
\end{center}

\begin{center}
\begin{figure}[ph]
\vspace*{-1.cm}
\hbox to 1.5cm
{\hspace*{3.cm}\hfil\mbox{$\tau_{\Xi_{bb}^{0}}$, ps}}
\vspace*{6.5cm} \hbox to 17.5cm {\hfil \mbox{$|\Psi^{dl} (0)|^2$,
GeV$^{3}$}} \vspace*{8.cm}\hspace*{2.5cm} \epsfxsize=12cm
\epsfbox{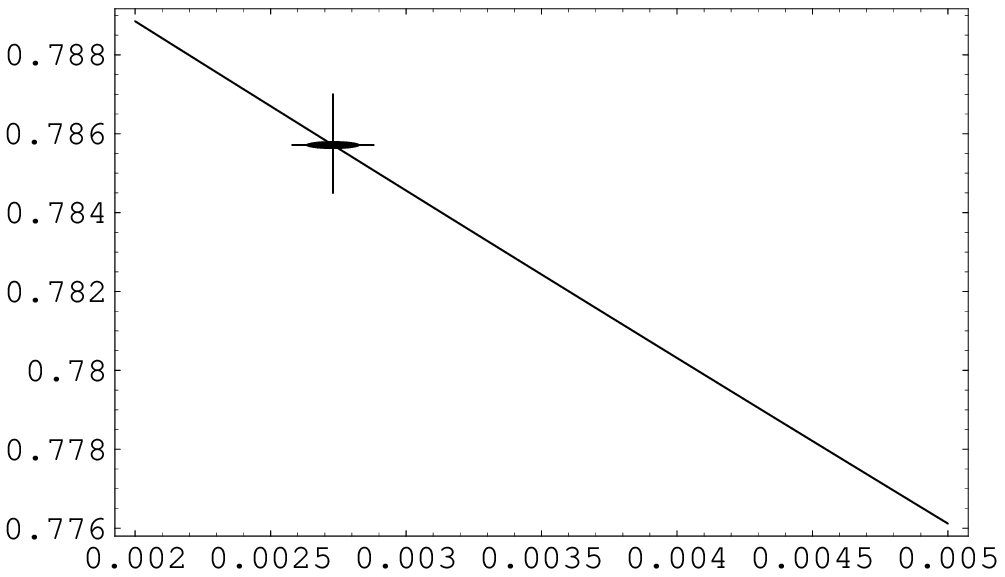} \vspace*{-15.cm}
\caption{The dependence of
$\Xi_{bb}^{0}$-baryon lifetime on the value of wavefunction of
light quark-diquark system at the origin $|\Psi^{dl} (0)|$.}
\label{bbu}
\end{figure}
\end{center}

\begin{center}
\begin{figure}[ph]
\vspace*{-1.cm}
\hbox to 1.5cm
{\hspace*{3.cm}\hfil\mbox{$\tau_{\Xi_{bb}^{-}}$, ps}}
\vspace*{6.5cm} \hbox to 17.5cm {\hfil \mbox{$|\Psi^{dl} (0)|^2$,
GeV$^{3}$}} \vspace*{8.cm}\hspace*{2.5cm} \epsfxsize=12cm
\epsfbox{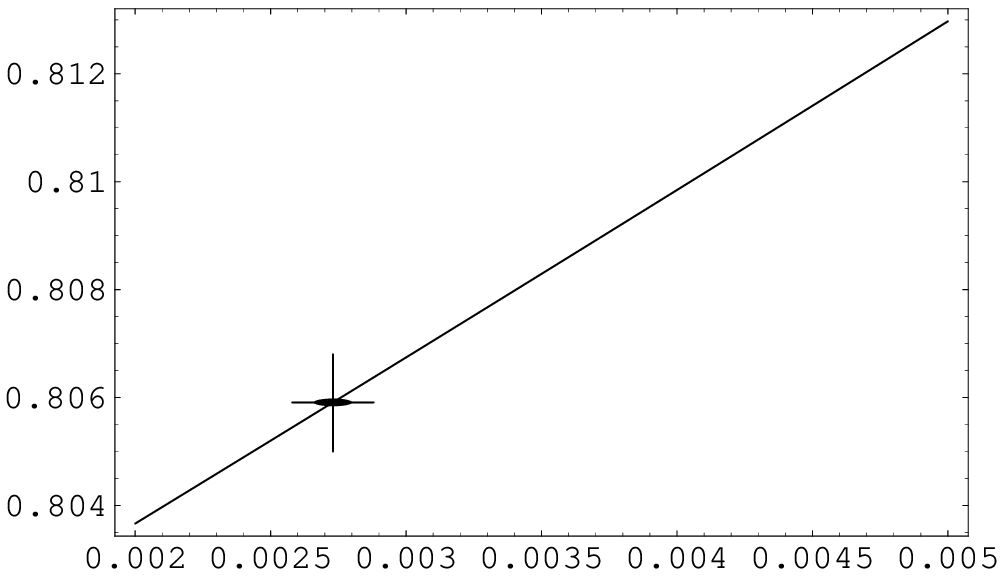} \vspace*{-15.cm}
\caption{The dependence of
$\Xi_{bb}^{-}$-baryon lifetime on the value of wavefunction of
light quark-diquark system at the origin $|\Psi^{dl} (0)|$.}
\label{bbd}
\end{figure}
\end{center}

\begin{center}
\begin{figure}[ph]
\vspace*{-1.cm}
\hbox to 1.5cm
{\hspace*{3.cm}\hfil\mbox{$\tau_{\Omega_{bb}}$, ps}}
\vspace*{6.5cm} \hbox to 17.5cm {\hfil \mbox{$|\Psi^{dl} (0)|^2$,
GeV$^{3}$}} \vspace*{8.cm}\hspace*{2.5cm} \epsfxsize=12cm
\epsfbox{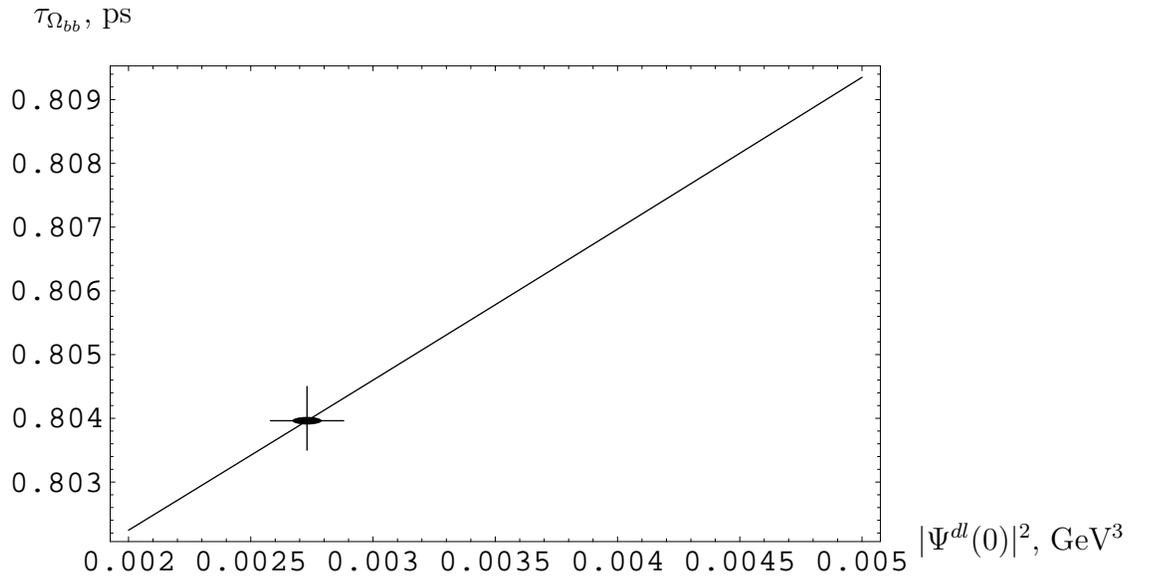} \vspace*{-15.cm}
\caption{The dependence of $\Omega_{bb}$-baryon lifetime on the value of
wavefunction of
light quark-diquark system at the origin $|\Psi^{dl} (0)|$.}
\label{bbs}
\end{figure}
\end{center}

In Figs. \ref{ccu}-\ref{bbs} we have shown the dependence of
baryons lifetimes from the values of light quark-diquark
wavefunctions at the origin. We see quite a different behaviour
withthe increase of $|\Psi^{dl} (0)|$-parameter.

%\vspace*{-6.cm}
Here, we would like to note, that in this paper we do not
give a detail discussion of nonspectator effects on the total
lifetimes and semileptonic branching ratios of doubly heavy
baryons and promise to fill this gap in one of our subsequent
papers \cite{Onishchenko1}.

Finally, concerning the uncertainties of the presented estimates,
we note that they are mainly determined by the following:

1) The $c$-quark mass is poorly known, but constrained by the fits
to the experimental data, discussed above, can lead to the
uncertainty $\frac{\delta\Gamma}{\Gamma}\approx 15\%$ in the case
of doubly charmed baryons and $\frac{\delta\Gamma}{\Gamma}\approx
10\%$ for the case of $bcq$ - baryons.

2) The uncertainties in the values of diquark and light quark-diquark
wavefunctions lead to $\frac{\delta\Gamma}{\Gamma}\approx 30\%$ in the case of
doubly charmed baryons and $\frac{\delta\Gamma}{\Gamma}\approx 15\%$ for
the $bcq$ - baryons.

Thus, the estimated uncertainty in predictions for the lifetimes
of doubly heavy baryons is close to $25\%$ in the case of $(bcq)$
- baryons, of order of $45\%$ in the case of doubly charmed
baryons and less then $5\%$ in the case of $(bbq)$ - baryons.
\vspace*{0.5cm}

\section{Conclusion}

In the present paper we have performed a detail investigation and
numerical estimates for the lifetimes of doubly heavy
baryons. The used approach is based on OPE expansion of total widths
for the corresponding hadrons, and it is combined with the formalism of
effective fields theories developed previously. In this way, we have
accounted for the both perturbative QCD and mass corrections to the
Wilson coefficients of operators. The nonspectator effects,
presented by Pauli interference and weak scattering, and their
influence on the total lifetimes are considered. The obtained
results show the significant role played by them in the
description of lifetimes of doubly heavy baryons.

The authors thank V.V.Kiselev for fruitful discussions and remarks concerning
the presentation of results.

This work is in part supported by the Russian Foundation for Basic
Research, grants 96-02-18216 and 96-15-96575. The work of
A.I.~Onishchenko was supported by International Center of
Fundamental Physics in Moscow and Soros Science Foundation.

\end{document}